\documentstyle[amstex,amssymb]{article}
\newcommand{\ZZ}{\bf Z}

\newcommand{\NN}{\bf N}

\newtheorem{theorem}{Theorem}
\newtheorem{lemma}{Lemma}[section]
\newtheorem{prop}[lemma]{Proposition}
\newtheorem{coro}{Corollary}
\newtheorem{definition}{Definition}
\begin{document}
\title{Uniform spectral properties of one-dimensional quasicrystals,\\
  I. Absence of eigenvalues} 
\author{David Damanik and Daniel Lenz
\\Fachbereich Mathematik\\Johann Wolfgang
Goethe-Universit\"at\\60054 Frankfurt/Main\\Germany}  
\maketitle
\begin{abstract}
We consider discrete one-dimensional Schr\"odinger operators with
Sturmian potentials. For a full-measure set of rotation numbers
including the Fibonacci case we prove absence of eigenvalues for all
elements in the hull.
\end{abstract}

\section{Introduction}
In this paper we consider discrete one-dimensional Schr\"odinger
operators in $l^2(\ZZ)$ with Sturmian potentials, namely,
\begin{equation}\label{hlt}
(H_{\lambda,\alpha,\theta}u)(n)=u(n+1)+u(n-1)+\lambda v_{\alpha,\theta}(n)u(n),
\end{equation}
where 
\begin{equation}\label{stpot}
v_{\alpha,\theta}(n)=\chi_{[1-\alpha,1)}(n\alpha +\theta \, mod \, 1),
\end{equation}
$\lambda \not= 0$, $\alpha \in (0,1)$ irrational and $\theta \in [0,1)$,
along with the corresponding difference equation 
\begin{equation}\label{eve}
H_{\lambda,\alpha,\theta} u=Eu.
\end{equation} 
The operator family (\ref{hlt}) describes a standard
one-dimensional quasicrystal model \cite{lp1,s} and has been studied in
many papers, e.g. \cite{bist,bit,d1,irt,k1,r}. Moreover, the operators
$H_{\lambda,\alpha,\theta}$ have attracted attention since they exhibit
spectral properties such as  zero-measure spectrum and purely singular
continuous spectral measures that seemingly hold for the entire family
in contrast to the almost Mathieu operator where similar properties were
shown to hold for a strict subclass of parameter values \cite{j,l1}.\\
\\
To put our present study into perspective, let us consider the class
of discrete one-dimensional Schr\"odinger operators with strictly
ergodic (i.e., minimal and uniquely ergodic), aperiodic potentials
taking finitely many values. Among these, potentials generated by
primitive substitutions and circle maps have received particular interest
\cite{bbg1,bbg2,bg1,bg2,d2,d3,dp1,dp2,h,hks,it,k1,s3,s4}. Sturmian
potentials (\ref{stpot}) are circle map potentials sharing some
crucial properties with potentials generated by primitive
substitutions. The general belief is that, as a rule, the spectrum has
zero Lebesgue measure and the spectral measures are purely singular
continuous. By results of Kotani \cite{k2} and Last-Simon \cite{ls},
absence of absolutely continuous spectrum holds in full generality, that
is, for every such family of operators and for every element of the
family. Zero-measure spectrum was proven for all Sturmian potentials by
Bellissard {\it et al.} \cite{bist} and for a large class of primitive
substitutions by Bovier-Ghez \cite{bg1,bg2}. Again, the zero-measure
property holds for all elements of the family since, by minimality, the
spectrum is constant over the hull. On the other hand, absence of point
spectrum has not yet been shown to hold in similar generality. Generic
absence of eigenvalues for certain models was proven in
\cite{bbg1,bist,bg1,bg2,d1,dp2,hks,s3}, while the works
\cite{d2,d3,dp1,k1} contain almost sure results. However, no uniform
result, that is, absence of eigenvalues for an entire such family, was
known yet. For the particular case of Sturmian potentials, generic
absence of eigenvalues is essentially due to Bellissard {\it et al.}
\cite{bist} (the paper does not state the result, see \cite{d1,hks} for
the result and proofs), whereas almost sure absence of eigenvalues was shown by
Kaminaga \cite{k1}, extending an argument of Delyon-Petritis \cite{dp1}
who had already obtained a partial result.\\
\\ 
Our purpose here is to prove the following theorem. 

\begin{theorem}\label{theo}
Suppose $\limsup a_n \not= 2$, where the $a_n$ are the
coefficients in the continued fraction expansion of $\alpha$. Then, for every
$\lambda$ and every $\theta$, the operator $H_{\lambda,\alpha,\theta}$
has empty point spectrum. 
\end{theorem}
{\it Remarks.}
\begin{enumerate}
\item The set of $\alpha$'s obeying the assumption of Theorem
  \ref{theo} has full Lebesgue measure \cite{khin}.
\item In particular, the Fibonacci case $\alpha = \frac{\sqrt{5}-1}{2}$
  is included as all the $a_n$ are equal to $1$ in this case.
\item Some works associate a slightly different family to the parameters
  $\lambda,\alpha$ \cite{h,hks} which is larger than the family
  parametrized by $\theta \in [0,1)$. The proof also works for the
  additional elements in that larger hull.
\item Our approach is based upon the two-block method \cite{g,s3} and
  yields additional information about stability properties that will be
  discussed within a more general context in \cite{dl2}.
\item Another key ingredient in our proof is an analogue to the hierarchical
  structures and the concept of (de-)composition in self-similar tilings
  \cite{gs}. This will be further exploited in \cite{dl1}.
\end{enumerate}
Thus, combining Theorem \ref{theo} with the results of Bellissard {\it et
  al.} \cite{bist}, the general picture proves to be correct for most
parameter values.

\begin{coro}
Suppose $\alpha$ obeys the assumption of Theorem \ref{theo}. Then, for
every $\lambda$ and every $\theta$, the operator
$H_{\lambda,\alpha,\theta}$ has purely singular continuous spectrum
supported on a Cantor set of zero Lebesgue measure.
\end{coro}
The organization of the paper is as follows. Section 2 recalls basic
properties of Sturmian potentials as well as the two-block
method. Hierarchical structures in Sturmian sequences are studied in
Section 3, while Section 4 provides a proof of Theorem \ref{theo}.

\section{Basic properties of Sturmian potentials}
Since our argument will be based upon the two-block method, our strategy
will be to exhibit appropriate squares adjacent to a certain site $i \in
\ZZ$. This
approach is entirely independent of the actual numerical values the
potential takes, and we can, without loss of generality, restrict our
attention to the particular case $\lambda = 1$. That is, we shall study
the combinatorial properties of the sequences $v_{\alpha,\theta}$. Consider 
such a sequence as a two-sided infinite word over the alphabet
$A \equiv \{0,1\}$. All these words share the property that their subword
complexity is minimal among the aperiodic sequences. Infinite words with
this property are called {\it Sturmian}. We shall recall some basic
facts about these words that will be used in Sections 3 and 4. The
material is taken from \cite{bist,b}. Moreover, as a general reference
on Sturmian words we want to mention the forthcoming monograph \cite{l2}.\\
\\
Fix $\alpha$ and consider its continued fraction expansion (for general
information on continued fractions see, e.g., \cite{khin}),
\begin{equation}
  \label{continuedfraction}
  \alpha = \cfrac{1}{a_1+ \cfrac{1}{a_2+ \cfrac{1}{a_3 + \cdots}}}
  \equiv [a_1,a_2,a_3,...]
\end{equation}
with uniquely determined $a_n \in \NN$. The associated rational
approximants $\frac{p_n}{q_n}$ obey 
$$p_0 = 0, \; p_1 = 1, \; p_n = a_n p_{n-1} + p_{n-2},$$
$$q_0 = 1, \; q_1 = a_1, \; q_n = a_n q_{n-1} + q_{n-2}.$$
Define the words $s_n$ over the alphabet $A$ by

\begin{equation}
  \label{recursive}
s_{-1} \equiv 1, \;\; s_0 \equiv 0, \;\; s_1 \equiv s_0^{a_1 - 1} s_{-1},
\;\; s_n \equiv s_{n-1}^{a_n} s_{n-2}, \; n \ge 2.
\end{equation}
In particular, the length of the word $s_n$ is equal to $q_n$, $n \ge 0$.
\begin{prop}\label{palin}
There exist palindromes $\pi_n$, $n \ge 2$, such that
\begin{equation}
  \label{even}
  s_{2n} = \pi_{2n}10,
\end{equation}
\begin{equation}
\label{odd}
  s_{2n+1} = \pi_{2n+1}01
\end{equation}
\end{prop}
{\it Proof.} See \cite{b}, where also the recursive relations obeyed by
the $\pi_n$ can be found.\hfill$\Box$\\
\\
By definition, for $n \ge 2$, $s_{n-1}$ is a prefix of $s_n$. Therefore,
the following (``right''-) limit exists in an obvious sense,
\begin{equation}
  \label{standard}
c_\alpha \equiv \lim_{n \rightarrow \infty} s_n.  
\end{equation}
Similarly, from (\ref{recursive}) we infer that, for $n \ge 1$,
$s_{2n-2}$ is a suffix of $s_{2n}$, and hence, the following (``left''-) limit
exists,
\begin{equation}
  \label{standleft}
d_\alpha \equiv \lim_{n \rightarrow \infty} s_{2n}.  
\end{equation}
The exceptional role played by the sequence $v_{\alpha,0}$ is
demonstrated by the following proposition.

\begin{prop}\label{vsymm}
$v_{\alpha,0}$ restricted to $\{1,2,3,\ldots\}$ coincides with $c_\alpha$,
$v_{\alpha,0}$ restricted to $\{\ldots,-2,-1,0\}$ coincides with $d_\alpha$.
\end{prop}
{\it Proof.} The first claim was shown by Bellissard {\it et al.} in
\cite{bist}. The second claim follows from the first combined with
Proposition \ref{palin} and the symmetry $v_{\alpha,0}(-k) =
v_{\alpha,0} (k-1)$, $k \ge 2$, also shown in \cite{bist}.\hfill$\Box$ 
\\
\\
For later use, we also want to note the following elementary formula.

\begin{prop}\label{wunderformel}
For each $n \ge 2$, $s_n s_{n+1}= s_{n+1} s_{n-1}^{a_n - 1} s_{n-2} s_{n-1}$.
\end{prop}
{\it Proof.} $s_n s_{n+1} = s_n s_n^{a_{n+1}} s_{n-1} = s_n^{a_{n+1}}
s_n s_{n-1} = s_n^{a_{n+1}} s_{n-1}^{a_n} s_{n-2} s_{n-1} = s_{n+1}
s_{n-1}^{a_n - 1} s_{n-2} s_{n-1}.$\hfill$\Box$\\
\\
The point is that the word $s_n s_{n+1}$ has $s_{n+1}$ as a prefix. Note
that the dependence of $a_n, p_n, q_n, s_n, \pi_n$ on $\alpha$ is 
left implicit. This, however, should not lead to any real confusion as
$\alpha$ will always be fixed within a local context.\\
\\
We now turn to the study of generalized eigenfunctions, i.e. solutions
of (\ref{eve}). Recall the standard reformulation of (\ref{eve}) in
terms of transfer matrices, namely, 
$$U(n+1)=T_{\lambda,\alpha,\theta}(n,E)U(n),$$
where $u$ is a solution of (\ref{eve}),
$$U(n) \equiv \left( \begin{array}{c} u(n)\\u(n-1) \end{array}\right),$$
and
$$T_{\lambda,\alpha,\theta}(n,E) \equiv \left(
\begin{array}{cc}
E-\lambda v_{\alpha,\theta}(n) & -1\\
1 & 0
\end{array}
\right).$$
Thus, the sequence $(u(n))_{n \in \ZZ}$ is determined by two consecutive
values, say $u(0)$ and $u(1)$, and all other values can be determined by
applying a matrix product of the form $M_{\lambda,\alpha,\theta}(n,E)
\equiv T_{\lambda,\alpha,\theta}(n,E) \times \cdots \times
T_{\lambda,\alpha,\theta}(1,E)$ to the vector $U(1)$ (for $n \ge 1$, the
case $n \le 0$ is similar). Note that $\det(M_{\lambda,\alpha,\theta}(n,E))=1$.
Hence, the characteristic equation of $M_{\lambda,\alpha,\theta}(n,E)$
takes the form 

\begin{equation}
  \label{char}
  M_{\lambda,\alpha,\theta}(n,E)^2 - tr(M_{\lambda,\alpha,\theta}(n,E))
  M_{\lambda,\alpha,\theta}(n,E) + I = 0.
\end{equation}
Now, the spectrum of $H_{\lambda,\alpha,\theta}$ is independent of
$\theta$ \cite{bist} and can thus be denoted by
$\Sigma_{\lambda,\alpha}$. 

\begin{prop}\label{tracebound}
For every $\lambda \not= 0$, there exists $C_\lambda$ such that, for
every irrational $\alpha$, every $E \in \Sigma_{\lambda,\alpha}$ and
every $n \in \NN$, we have $| tr(M_{\lambda,\alpha,0}(q_n,E) | < C_\lambda$.
\end{prop}
{\it Proof.} See \cite{bist}.\hfill$\Box$\\
\\
Before we formulate our basic criterion for absence of eigenvalues, we
introduce the following two concepts. Let $v$ be a two-sided sequence
over $A$ (think of $v=v_{\alpha,\theta}$). A subword $x=x_1 \ldots x_l$
of $v$ is called {\it adjacent to $i \in \ZZ$} if $v_i \ldots
v_{i+l-1} = x$ or $v_{i-l+1} \ldots v_i = x$. Two finite words $x,y$
having the same length are called {\it conjugate} if $x$ is a subword
of $yy$. This notion is easily seen to induce an equivalence relation on
$A^l$, for any fixed $l$. Intuitively, the equivalence class of a word $x=x_1
\ldots x_l$ is given by the collection of all cyclic permutations $x_{j+1}
\ldots x_l x_1 \ldots x_j$ of $x$.
 
\begin{lemma}\label{zweiblock}
Suppose $\alpha,\theta$ are such that $v_{\alpha,\theta}$ has
infinitely many squares $u_k u_k$ adjacent to some $i \in \ZZ$, where $u_k$ is
conjugate to some $s_{n_k}$. Then, for every $\lambda$, $\sigma_{pp}
(H_{\lambda, \alpha,\theta}) = \emptyset$.
\end{lemma}
{\it Proof.} In case $i=1$, the assertion follows by standard arguments
\cite{g,s3} from (\ref{char}) together with Proposition
\ref{tracebound}. The general case can be reduced to this case by a
suitably chosen shift of the sequence $v_{\alpha,\theta}$, which, of
course, leaves the spectral type of the associated operators
$H_{\lambda,\alpha,\theta}$ invariant.\hfill$\Box$\\
\\We close this section by introducing the shift operator $T$ on
functions on $\ZZ$, i.e. 
$$ (T f) (n) \equiv f(n+1)$$
for arbitrary functions $f$ on $\ZZ$.

\section{The partition lemma} Consider for a fixed  $\alpha$ the family (in $\theta$) of all
the sequences of the form $(v_{\alpha,\theta}(k))_{k\in \ZZ}$. It is
well known that in each sequence of the family the same words
occur. Moreover any of these words occurs with a fixed frequency greater than
zero which is independent of the sequence (cf. Appendix of
\cite{h}). Thus, from a measure theoretical point of view, the family 
$\{(v_{\alpha,\theta}(k))_{k\in \ZZ}\,|\,\theta \in [0,1)\}$ behaves
very uniformly. However, to
prove the absence of eigenvalues for all $\theta$ we need a kind of
uniform topological structure. In this section we provide this structure
by showing that each sequence $(v_{\alpha,\theta}(k))_{k\in \ZZ}$ can
be decomposed into blocks of the form $s_n$ and $s_{n-1}$
for all $n\in \NN_0$. Here  we denote by $\NN_0$ the set of all integers
together with $0$, i.e. $\NN_0=\{0,1,2,\ldots\}$.

\begin{definition}\label{n-partition}Let $n\in \NN_0$ be given. An 
    $(n,\alpha)$-partition of  a sequence $(f_k)_{k\in\ZZ}$ with $f_k\in\{0,1\}$   is a sequence $(I_j, z_j)_{j\in\ZZ}$ of pairs
    $I_j=\{d_j,d_j+1,\ldots,d_{j+1}-1\}\subset \ZZ$ and   $z_j\in \{s_n,
s_{n-1}\}$  with $0\in I_0$ s.t.
$$f_{d_j}f_{d_j +1}...f_{d_{j+1}-1}=z_j$$
for all $j\in \ZZ$.\\
An $(n,\alpha)$-partition of  a function $f:\ZZ \longrightarrow \{ 0,1 \}$
is an $(n,\alpha)$-partition of the sequence $(f(n))_{n\in \ZZ}$.\\
The $z_j$ are referred to as blocks in the $n$-partition or more specifically
as blocks of the form $s_n$ if $z_j=s_n$, and as blocks of the form
$s_{n-1}$ if $z_j=s_{n-1}$.  The $I_j$ are referred to as positions of
the blocks $z_j$.             
\end{definition}
In the sequel we will sometimes suppress the dependence on $\alpha$ if
it is clear from the context to which $\alpha$ we refer. In particular,
we will write $n$-partition instead of $(n,\alpha)$-partition.\\
\\
{\it Remarks.}\begin{enumerate}
\item One can think of an $n$-partition  as a tiling
of $\ZZ$ by $s_n$ and $s_{n-1}$ generating $f$. So $f$ is composed out
of the  blocks $z_j$ at the positions $I_j$. 
\item Our notion of $n$-partition is analogous to the notion of
$n$th composition used in the study of self similar tilings
 \cite{lp}. This notion has been used by Hof in \cite{h}. There he studies the Lyapunov exponent and the integrated density of states of discrete Schr\"odinger operators with a potential generated by
a primitive substitution. 
\end{enumerate}
As $s_0=0$ and $s_{-1}=1$ every $f:\ZZ \longrightarrow \{ 0,1 \}$ admits a
$0$-partition. But for a general  $f$ there does not necessarily
exist an $n$-partition for $n>0$. However, it is crucial to our
analysis of the eigenvalue problem for $H_{\lambda,\alpha,\theta}$ that for
the sequences $v_{\alpha,\theta}$ there do exist unique $n$-partitions
for all $n\in \NN_0$. This is the content of the next lemma.

\begin{lemma}\label{partition-lemma}
\begin{itemize}
\item[(i)] For every $n\in \NN_0$, there exists a unique $n$-partition of
  $v_{\alpha,0}$.
\item[(ii)] For every $n\in \NN_0$ and every $\theta \in [0,1)$, there exists
  a unique $n$-partition of $v_{\alpha,\theta}$.
\end{itemize}
\end{lemma}
{\it Proof.}  Let $\alpha=[a_1,a_2,...]$ be the continued fraction
expansion of $\alpha$ (cf. equation \ref{continuedfraction}).\\
(i) {\it Existence:} Set $v \equiv v_{\alpha,0}$. We show that there are
$n$-partitions of $(v(k))_{k\geq 1}$ and of
$(v(k))_{k\leq 0}$. Here an $n$-partition of a one sided
sequence is defined in the obvious way. By (\ref{recursive}),
(\ref{standard}) and Proposition \ref{vsymm},
 it is clear that there exists an
$n$-partition of  $ (v(k))_{k\geq 1}$  for all $n$.\\
The existence of an $n$-partition  for $ (v(k))_{k\leq 0}$
follows similarly by (\ref{recursive}),
(\ref{standleft}) and Proposition \ref{vsymm}.   \\
{\it Uniqueness:} This follows by induction: As $s_0=0, s_{-1}=1$ uniqueness
is clear for $n=0$. By (\ref{recursive}), every $(n+1)$-partition  gives
rise to an $n$-partition  and the positions of 
the $s_{n+1}$ in the $(n+1)$-partition are determined by the positions of
$s_{n-1}$  in the $n$-partition. Thus, uniqueness of the $n$-partition
implies uniqueness of the $(n+1)$-partition.\\
\\
(ii) Fix $\theta \in [0,1)$.  As $\alpha$ is irrational there exists a
sequence $(n_k)_{k\in  \NN}$, $n_k\in \NN$, s.t. the sequence 
$(T^{n_k} v_{\alpha, 0})_{k\in \NN}$ converges to
$ v_{\alpha, \theta}$ in the product topology on $\{0,1\}^{\ZZ}$ for $k\to \infty$. By (i), it is clear that the
$ T^{n_k} v_{ \alpha, 0}$ admit unique $n$-partitions  for all
$n\in \NN_0$. To use this to conclude (ii) we introduce the following notion of
convergence:\\
\\
Let $f_k,k\in \NN$, and $f$ be functions for which there exist unique
$n$-partitions  denoted by $(I_j^k, z_j^k)$ and $(I_j,z_j)$,
respectively. We say that the $f_k$ converge to $f$ in the $n$-sense for
$k\to \infty$ if for all $C>0$ there exists a $k_0$ s.t. for $k\geq k_0$ 
$$(I_j^k, z_j^k)=(I_j,z_j) \mbox{ for all}\;\: I_j\subset (-C,C)   $$
holds.
\\
Clearly, (ii) follows if we prove the following claim.\\
\\
{\it Claim.}  For each $n$ there exists a unique $n$-partition of
$v_{\alpha,\theta}$ and  the sequence $(T^{n_k}v_{\alpha,0})_{k\in \NN_0}$ converges to
$v_{\alpha,\theta}$ in the $n$-sense for $k\to \infty$.\\
\\
{\it Proof of the claim.} This will be done by induction. We will
consider two cases.\\[1mm]
{\it Case 1:} $a_1=1$.\\
As $s_{-1}=s_{1}=1$ and $s_0=0$, the cases $n=0$ and $n=1$ are clear. 
 So suppose the statement is true for $n\geq 1$ fixed. Let
$(I_j,z_j)$ be the $n$-partition of $v_{\alpha,\theta}$. By
$s_{n+1}=s_n^{a_{n+1}} s_{n-1}$ (cf. equation \ref{recursive}), the
existence of an $(n+1)$-partition of $v_{\alpha,\theta}$ will follow if
we show that to the left of each block of the form $s_{n-1}$ in the
$n$-partition of  $v_{\alpha,\theta}$ there are at least $a_{n+1}$
blocks $s_n$. That is, we have to show that $z_j=s_{n-1}$ for 
$j\in \ZZ$ implies $z_k=s_n$ for $k= j - a_{n+1}, ..., j-1$.
 As $T^{n_k}v_{\alpha,0}$ admits a
 unique $n$-partition for each $n\in \NN_0$, there are at least
 $a_{n+1}$ blocks $s_n$ to the left of each block of the form $s_{n-1}$
 in the $n$-partition of $T^{n_k}v_{\alpha,0}$. As the
$T^{n_k}v_{\alpha,0}$ converge to $v_{\alpha,\theta}$ in the
$n$-sense, the corresponding statement is true for
$v_{\alpha,\theta}$. This gives the existence of an $(n+1)$-partition of
$v_{\alpha,\theta}$. The uniqueness of the $(n+1)$-partition follows from
the uniqueness of the $n$-partition as in $(i)$. As the blocks $s_{n+1}$
in the $(n+1)$-partition of $v_{\alpha,\theta}$ arise from blocks
$s_n^{a_{n+1}} s_{n-1}$ in the $n$-partition of $v_{\alpha,\theta}$, it
is clear that the convergence of $T^{n_k}v_{\alpha,0}$ to
$v_{\alpha,\theta}$ in the $n$-sense implies the convergence in the
$(n+1)$-sense. This proves the claim in Case 1.\\[1mm]
{\it Case 2:} $a_1>1$.\\
As $s_{-1}=1$ and $s_0=0$, the case $n=0$  is  clear. 
So fix $n\geq 0$. If $n>0$ we can continue  exactly as in Case 1. If
$n=0$ we can continue as in Case 1 after replacing $a_{n+1}=a_1$ by
$a_{1}-1\geq1$. This proves the claim in Case 2. \\
The proof of the lemma is therefore finished.\hfill $\Box$
\begin{coro}\label{potenzen} Let, for $n\in \NN$,  $(I^n_j,z^n_j)$ be
  the $n$-partition of $v_{\alpha,\theta}$. If $z^n_j=s_{n-1}$, then
  $z^n_{j-1}=z^n_{j+1}=s_n$. If $z^n_j=s_n$, then there is an interval
  $I=\{d,d+1,\ldots,d+l-1\}\subset \ZZ$ of length $l\in\{a_{n+1},a_{n+1}
  +1\}$ with $j\in I$ and $z^n_i=s_n$ for all $i\in I$ and
  $z^n_{d-1}=z^n_{d+l}=s_{n-1}$. 
\end{coro}
{\it Proof.} By the existence part of the partition lemma, there exists an
$(n+1)$-partition of $v_{\alpha,\theta}$. By the uniqueness part of the
partition lemma and the formula $s_{n+1}= s_n^{a_{n+1}} s_{n-1}$, all
the blocks of the form $s_{n-1}$ in the $n$-partition of
$v_{\alpha,\theta}$ arise from blocks of the form  $s_{n+1}$ in the
$(n+1)$-partition. This shows that there is no $j\in \ZZ$ with
$z_j^n=z_{j+1}^n=s_{n-1}$ and that there are at least $a_{n+1}$ blocks of
the form $s_n$ between two blocks of the form $s_{n-1}$. That there are
at most $a_{n+1}+1$ such blocks follows, as there are not two adjacent
blocks of the form $s_n$ in the $(n+1)$-partition. This proves the
corollary.\hfill $\Box$\\ 
\\{\it Remarks.}\begin{enumerate}
\item  Define $\Omega$ to be the set of accumulation points of 
  translates of $v_{\alpha,0}$ with respect to pointwise convergence, that is,
$$\Omega(\alpha) \equiv \{ \omega \in A^{\ZZ} \; | \; \omega = \lim T^{n_i}
v_{\alpha,0}, \; n_i \rightarrow \infty\}.$$
Then the method of the previous lemma can easily be adopted to prove the
existence of a unique $n$-partition  for all $\omega\in
\Omega(\alpha)$. We refer the reader to  \cite{dl2} for a  discussion of
the relationship between $\Omega(\alpha) $ and the set
$\{v_{\alpha,\theta}\;|\; \theta \in [0,1)\}.$
\item The above lemma and a careful analysis of $v_{\alpha,0}$ allow to
show that the blocks $s_n$ and $s_{n-1}$ in the $n$-partition of an
$\omega \in \Omega(\alpha)$  occur with fixed frequency, see
\cite{dl1} for details. This can 
be used together with Theorem 1 of \cite{gh} to show that indeed every word
that occurs in some $\omega_0\in \Omega(\alpha)$ occurs in every $\omega\in \Omega(\alpha)$ with a fixed frequency greater than zero. Thus, the
above partition lemma is indeed a stronger result than the results about
the frequencies of words mentioned at the beginning of this section.
\end{enumerate}
\section{Absence of eigenvalues}

Let $\alpha=[a_1,a_2,...]$ be the continued fraction
expansion of $\alpha$ (cf. equation \ref{continuedfraction}).\\
\\
The proof of Theorem \ref{theo} will be split into two parts. 

\begin{prop}\label{drei} If $\limsup_{n\to \infty} a_n\geq 3$, then, for
  every $\theta$ and every $\lambda$, the operator $H_{\lambda,\alpha,
    \theta}$ has no eigenvalues. 
\end{prop}

\begin{prop}\label{fib} If $\limsup_{n\to \infty} a_n=1$, then, for every
  $\theta$ and every $\lambda$, the operator $H_{\lambda,\alpha, \theta}$
  has no eigenvalues. 
\end{prop}
{\it Remark.} In fact the proofs yield the absence of eigenvalues for all
potentials in the respective hulls
$$\Omega(\lambda,\alpha):=\{\lambda \omega: \omega\in \Omega(\alpha)\}.$$
\\{\it Proof of Proposition \ref{drei}.} Fix $\theta\in [0,1)$. We will
exclude eigenvalues of $H_{\lambda,\alpha,\theta}$ using  Lemma
\ref{zweiblock}, that is, by exhibiting many appropriate squares in
$v_{\alpha,\theta}$ 
at zero. By the partition lemma \ref{partition-lemma}, there is for each
$n\in \NN_0$ an $n$-partition 
$$ ((I_j^n,z_j^n))_{j\in \ZZ}, \;I_j^n=\{d_j^n,\ldots, d_{j+1}^n-1\},
\;z^n_j\in\{s_n,s_{n-1}\}$$
of $v_{\alpha,\theta}$. We will consider two cases.\\[1mm]
{\it Case 1}: There are infinitely many $n\in \NN_0$ with
$z^n_0=s_{n-1}$.\\
Consider such an $n$ with $n\geq 4$.  Corollary \ref{potenzen} yields
$z_1^n=s_n$. As $s_{n-1}$ is a prefix of $s_n$ and $z_2^n\in
\{s_n,s_{n-1}\}$ we have 
$$z_0^n z_1^n z_2^n=s_{n-1} s_n s_{n-1} w$$
 with a suitable word $w$. Using $s_n= s_{n-1}^{a_n} s_{n-2}$
 (cf. equation \ref{recursive}), we get
$$ z_0^n z_1^n z_2^n=s_{n-1} s_{n-1}^{a_n} s_{n-2} s_{n-1} w .$$
By $s_{n-2} s_{n-1}=s_{n-1} v$ with a suitable $v$ (cf. Proposition \ref{wunderformel}), we finally arrive at
$$ z_0^n z_1^n z_2^n=s_{n-1} s_{n-1}^{a_n} s_{n-1} v w.$$
This implies
$$v_{\alpha,\theta}(d_0^n) v_{\alpha,\theta}(d_0^n+1)... v_{\alpha,\theta}(d_0^n+ 3 |s_{n-1}| -1)= s_{n-1}   s_{n-1}  s_{n-1}
,$$
where $0\in \{d_0^n,..., d_0^n +  |s_{n-1}| -1)$. Thus, there exists a square
$x x$ with  $x$ being a cyclic permutation of $s_{n-1}$ to the right of
zero. Since this is true for infinitely many $n$, we can use Lemma
\ref{zweiblock} to exclude eigenvalues.\\[1mm]
{\it Case 2}: There is an $n_0\in \NN_0$ s.t. $z_0^n=s_n $ for all
$n\geq n_0$\\ 
By $\limsup_{n\to \infty} a_n\geq 3$, there are infinitely many $n\geq
n_0$ s.t. $a_n\geq 3$. Fix such an $n$. As $a_n\geq 3$ and $z_0^n=s_n$,
there are three cases by Corollary \ref{potenzen}.\\
\\{\it Subcase 1}: $z_0^n=z_1^n=z_2^n=s_n$.\\
In this case we have $z_0^n z_1^n z_2^n=s_n s_n s_n$.\\[1mm]
{\it Subcase 2}:  $z_0^n=z_{-1}^n=z_{-2}^n=s_n$.\\
In this case we have $ z_{-2}^n z_{-1}^n z_0^n=s_n s_n s_n$.\\[1mm]
{\it Subcase 3}: $z_{-1}^n=z_0^n=z_1^n= s_n$, $ z_{-2}^n=z_2^n=s_{n-1}$.\\
Calculating as in Case 1 we get $z_0^n z_1^n z_2^n z_3^n=s_n s_n s_n w$
with a suitable word $w$.\\
\\
Thus, in all subcases we can conclude as in Case 1 that there exists a
square $x x$, with $x$ being a cyclic permutation of $s_n$  either to the
left or to the right of zero. Since this applies to infinitely many $n$, we
can use Lemma \ref{zweiblock} to exclude eigenvalues in Case 2 as
well. This proves the proposition.\hfill $\Box$\\
\\
{\it Proof of Proposition \ref{fib}.} The proof is similar to the proof of
Proposition \ref{drei}. So fix $\theta\in [0,1)$. By  Lemma
\ref{partition-lemma}, there exists a unique $n$-partition
$ ((I_j^n,z_j^n))_{j\in \ZZ}$ of  $ v_{\alpha,\theta}.$  
Again we will consider two cases.\\[1mm]
{\it Case 1}: There are infinitely many $n\in \NN_0$ with $z^n_0=s_{n-1}$.\\
This case can be treated as in the proof of Theorem \ref{drei}.\\[1mm]
{\it Case 2}: There is an $n_0\in \NN_0$ s.t. $z_0^n=s_n $ for all $n\geq
n_0$.\\
By $\limsup_{n\to \infty} a_n=1$, there is an $n_1\in \NN_0$ s.t. $a_n=1$ for all $n\geq
n_1$. Let $c:=\max\{n_0,n_1\}$.\\
As $z_0^n=s_n$ and $s_{n+1}=s_n s_{n-1}$ for all $n\geq c$, it follows
easily by induction that
$$d_0^n=d_0^c \;\:\mbox{for all} \;\:n \geq c.$$
Moreover  $z_0^n=s_n$ and $s_{n+1}=s_n s_{n-1}$ imply $z_1^n=s_{n-1} $
for all $n\geq c$ and this in turn implies $z_2^n=s_n$ for all $n\geq
c$. We therefore have
$$ z_0^n z_1^n z_2^n=s_n s_{n-1} s_n= s_n s_n w,$$
where we used $s_{n-1} s_n= s_n w$ with a suitable word $w$
(cf. Proposition  \ref{wunderformel}). Putting all
this together, we see
$$ v_{\alpha,\theta}(d_0^c)...v_{\alpha,\theta}(d_0^c+ 2|s_n| -1)= s_n s_n \;\:\mbox{for all} \;\:n \geq c.$$
Again, an application of Lemma \ref{zweiblock} yields the desired
absence of eigenvalues, concluding the proof. \hfill $\Box$ 
\\
\\
{\it Acknowledgment.} One of us (D. L.) gratefully acknowledges
financial support from Studienstiftung des deutschen Volkes
(Doktorandenstipendium).

\end{document}